\documentclass[twocolumn]{autart} 
\usepackage{IEEEtrantools}

\usepackage{mathptmx} 
\usepackage{times} 
\usepackage{amsmath} 
\usepackage{amssymb}  

\usepackage[noadjust]{cite}
\newcommand{\abs}[1]{\left\lvert#1\right\rvert}
\usepackage{nicefrac}
\usepackage{graphicx}          
\usepackage[dvips]{epsfig}    
\usepackage{mathtools}
\mathtoolsset{showonlyrefs}
\newcommand{\defeq}{\coloneqq}
\newcommand{\eg}{\emph{e.g.}}
\newcommand{\ie}{\emph{i.e.}}
\usepackage{pgfplots}
\newlength\figureheight
\newlength\figurewidth
\begin{document}

\begin{frontmatter}

\title{ARX modeling of unstable linear systems\thanksref{footnoteinfo}} 

\thanks[footnoteinfo]{This work was supported by the European Research Council under
the advanced grant LEARN, contract 267381 and by the Swedish
Research Council under contract 621-2009-4017.}

\author[KTH]{Miguel Galrinho}\ead{galrinho@kth.se},   
\author[KTH]{Niklas Everitt}\ead{neveritt@kth.se},     
\author[KTH]{H{\aa}kan Hjalmarsson}\ead{hjalmars@kth.se}  

\address[KTH]{ACCESS Linnaeus Center, School of Electrical Engineering, KTH Royal Institute of Technology, Sweden}  


\begin{abstract}                          
High-order ARX models can be used to approximate a quite general class of linear systems in a parametric model structure, and well-established methods can then be used to retrieve the true plant and noise models from the ARX polynomials.
However, this commonly used approach is only valid when the plant is stable or if the unstable poles are shared with the true noise model. 
In this contribution, we generalize this approach to allow the unstable poles not to be shared, by introducing modifications to correctly retrieve the noise model and noise variance.
\end{abstract}

\end{frontmatter}

\section{Introduction}

The prediction error method (PEM) is a benchmark in system identification to obtain models for linear systems, since it provides asymptotically efficient estimates if the chosen model orders are correct~\cite{ljung99}. 
The drawback with PEM is that, in general, it requires solving a non-convex optimization problem, and may not attain the global minimum. 
However, for some particular model structures, minimizing the cost function of PEM is a linear regression problem. This is the case of autoregressive exogenous (ARX) models.

Despite the usefulness of ARX models due to the simplicity of estimation, they provide limited flexibility, not allowing the plant and the noise model to be parametrized independently. 
In this sense, Box-Jenkins (BJ) models are a more encompassing choice, as the plant and noise model are rational transfer functions parametrized independently. 
However, estimating BJ models with PEM suffers the drawbacks of solving a non-convex optimization problem.

Another advantage of ARX models is that, if the polynomial orders are allowed to increase, they can approximate a BJ model arbitrarily well. 
The limitation is that, as the number of parameters increases, so does the variance of the estimated model. 
Nevertheless, high-order ARX models are still useful: \eg, they can be used as an intermediate step to obtain a parsimonious model~\cite{wahlberg89,zhu11,hjalmarsson12,galrinho14}. 

If the plant is stable or if it shares all the unstable poles with the true noise model, it is well known that the plant and noise model can be recovered from the ARX polynomials~\cite{ljung&wahlberg92}. 
In this paper, we generalize these results to unstable plants, where some of the unstable poles are not shared by the noise model.
Although this is a non-standard case, because PEM with a correctly parametrized model has an unstable predictor, it makes sense: 
for example, if the noise model is used to model a sensor,
restricting this model to contain eventual unstable plant dynamics is unreasonable from a physical perspective.
To apply PEM in such a case, methods have been proposed to deal with unstable predictors~\cite{hjalmarsson99,forssell00}.

In this paper, we consider systems that encompass BJ structures, but also allow for shared poles between the plant and noise model.
Our contributions are the following. 
First, we derive the ARX polynomials obtained asymptotically when the plant has unstable poles not shared with the noise model.
Second, we observe that appropriate corrections are required to obtain consistent estimates of the noise model and the noise variance, and illustrate how these can be applied. 
Third, although the variance of the estimated model increases with the number of estimated parameters, we observe that the variance of the unstable poles remains small.

\section{Problem Statement}
Consider that data is generated by
\begin{equation}
y_t = G(q) u_t + H(q) e_t ,
\label{eq:truemodel}
\end{equation}
where $u_t$ is the plant input, $e_t$ is Gaussian white noise with variance $\lambda_e$, $y_t$ is the output, and $G(q)$ and $H(q)$ are the true plant and noise models, respectively, which are rational transfer functions in the delay operator $q^{-1}$, given by
\begin{IEEEeqnarray}{rClrCl}
G(q) & \defeq & \frac{L(q)}{\Gamma(q)F(q)} , \qquad & H(q) & \defeq & \frac{C(q)}{\Gamma(q)D(q)},
\label{eq:sys}
\end{IEEEeqnarray}
and
\begin{IEEEeqnarray*}{rCrl}
L(q)      & \defeq &     l_1 q^{-1} +& \dotsc + l_{m_l} q^{-m_l} , \\
\Gamma(q) & \defeq & 1 + \gamma_1 q^{-1} +& \dotsc + \gamma_{m_\gamma} q^{-m_\gamma} , \\
F(q)      & \defeq & 1 + f_1 q^{-1} +& \dotsc + f_{m_f} q^{-m_f} , \\
C(q)      & \defeq & 1 + c_1 q^{-1} +& \dotsc + c_{m_c} q^{-m_c} , \\
D(q)      & \defeq & 1 + d_1 q^{-1} +& \dotsc + f_{m_d} q^{-m_d} ,
\end{IEEEeqnarray*}
where $m_l$, $m_\gamma$, $m_f$, $m_c$, and $m_d$ are finite positive integers. 
Note that we are considering a more general class of models than BJ, which corresponds to $\Gamma(q) = 1$. 
Although the novelty of the results derived in this paper concerns the case where $G(q)$ and $H(q)$ are parametrized independently, 
so that $H(q)$ does not share eventual unstable poles of $G(q)$, we include for completeness the possibility that some poles are shared through the polynomial $\Gamma(q)$.
Since the possibility of $G(q)$ and $H(q)$ sharing poles is already included through $\Gamma(q)$, we assume that $F(q)$ and $D(q)$ are co-prime (i.e., they do not share common factors).

We impose that $C(q)$ and $D(q)$ are stable polynomials (\ie, all the roots lie inside the unit circle) and that $F(q)$ does not contain roots on the unit circle.
Since $\Gamma(q)$ and $F(q)$ are not required to be stable, we consider that the data is obtained with stabilizing feedback,
%
\begin{equation}
\left.
\begin{IEEEeqnarraybox}[][c]{rCrCr}
y_t & = & G(q)  S(q) r_t  & + & H(q) S(q)  e_t, 
\\
u_t & = & S(q) r_t & - & K(q)  H(q) S(q)  e_t,
\end{IEEEeqnarraybox}
\right.
\label{eq:syscl}
\end{equation}
where $r_t$ is a known external reference uncorrelated with $e_t$, $S(q) = [1+K(q)G(q)]^{-1}$ is the sensitivity function,
and $K(q)$ is a stabilizing regulator.

Consider also the ARX model
\begin{IEEEeqnarray}{rCl}
A(q) y_t & = & B(q) u_t + e_t ,
\label{eq:arxmodel}
\end{IEEEeqnarray}
with infinite order polynomials
\begin{IEEEeqnarray*}{rCl+rCl}
A(q) & = & 1+\sum_{k=1}^\infty a_k q^{-k} , &  B(q) & = & \sum_{k=1}^\infty b_k q^{-k} .
\end{IEEEeqnarray*}
Using a quadratic cost, the PEM estimate of the ARX model minimizes the cost function
\begin{IEEEeqnarray}{rCl}
J & \defeq & \mathbb{E} \left[ A y_t - B u_t \right]^2 ,
\label{eq:JtimeOld}
\end{IEEEeqnarray}
as the number of samples tend to infinity (here, the argument $q$ was dropped for notational simplicity).
Because the data are generated by~\eqref{eq:syscl}, the cost function can be expressed as
\begin{IEEEeqnarray}{rCl}
J & = & \mathbb{E} \left[ \left(A G - B \right) S r_t + \left(A + K B \right) H S e_t \right]^2 .
\label{eq:Jtime}
\end{IEEEeqnarray}
Then, the global minimizers of~\eqref{eq:Jtime}, $\bar{A}(q)$ and $\bar{B}(q)$, can be related to $G(q)$ and $H(q)$.

It is well known that, with the additional assumption that $F(q)$ is stable, \eqref{eq:Jtime} is minimized by~\cite{ljung&wahlberg92}
\begin{equation}
\left.
\begin{IEEEeqnarraybox}[][c]{rClCl}
\bar{A}(q) & = & \frac{1}{H(q)} & = & \frac{\Gamma(q)D(q)}{C(q)}, \\ 
\bar{B}(q) & = & \frac{G(q)}{H(q)} & = & \frac{D(q)}{C(q)}\frac{L(q)}{F(q)},
\end{IEEEeqnarraybox}
\right.
\label{eq:ABstar_stable_minphase} 
\end{equation}
with minimum 
\begin{IEEEeqnarray}{rCl}
J^\star & = & \lambda_e.
\label{eq:Jstar_stable_minphase}
\end{IEEEeqnarray}
Thus, an infinite order ARX model can be used to asymptotically recover $G(q)$, $H(q)$, and the noise variance $\lambda_e$.

In this paper, we seek the minimizers $\bar{A}(q)$ and $\bar{B}(q)$ of~\eqref{eq:Jtime} in the case of more general unstable plants, when $F(q)$ is also allowed to be unstable. In this case, the unstable poles of the plant are not shared with the noise model. 

\section{ARX Minimizer of Unstable BJ Model}
\label{sec:arxmin}

First, we introduce a result that will be used to derive our main result.
\begin{prop}
\label{thm:firstresult}
Let $Z(q)$ and its inverse be power series in $q^{-1}$ that are analytic outside and on the unit circle, and such that $Z(\infty)=1$. 
Let $X(q)$---also a power series in $q^{-1}$ satisfying $X(\infty)=1$---be the argument of the cost function
\begin{IEEEeqnarray}{rCl}
J & = & \frac{1}{2\pi} \int_{-\pi}^{\pi} \abs{X(e^{i\omega})}^2 \abs{Z(e^{i\omega})}^2 \text{ d}\omega .
\label{eq:Jthm1}
\end{IEEEeqnarray}
Then, \eqref{eq:Jthm1} has the unique minimizer $\bar{X}(q) = Z^{-1}(q)$ and minimum $J^\star = 1$.
\end{prop}

\begin{pf}
This is a standard result, see \emph{e.g.} Problem 3G.3 in \cite{ljung99}. A proof is included for completeness.
The product $X(q)Z(q)$ can be expanded as a polynomial,
\begin{IEEEeqnarray}{rCl}
X(q)Z(q) & = & \sum_{i=0}^{\infty} g_i q^{-i} ,
\label{eq:XZ}
\end{IEEEeqnarray}
where $g_0=1$, since ${X(\infty)=Z(\infty)=1}$.
Using Parseval's identity on~\eqref{eq:Jthm1} together with~\eqref{eq:XZ} yields
\begin{IEEEeqnarray*}{rCl}
J & = & 1 + \sum_{i=1}^{\infty} \abs{g_i}^2 \geq 1 .
\end{IEEEeqnarray*}
The minimum $J^*=1$ is obtained for ${\bar{X}(q) = Z^{-1}(q)}$, since $Z(q)$ is inversely stable. 
Because the inverse is unique, the minimum will not be attained for any other $X(q)$, since then at least one 
$g_i \neq 0, i > 0$. $\hfill \qed$
\end{pf}

Before stating the main result of the paper, we introduce the following definitions. 
Consider the factorization
\begin{IEEEeqnarray}{rCl}
F(q) & = & F_s(q) F_a(q),
\label{eq:FsFa}
\end{IEEEeqnarray}
where $F_s(q)$ and $F_a(q)$ contain the stable (magnitude less than one) and anti-stable (magnitude larger than one) roots of $F(q)$, respectively.
Also, we define $F_a^*(q)$ as the polynomial with the roots of $F_a(q)$ mirrored inside the unit circle, \ie, $F_a^*(q)  =  \prod_{k=1}^{n_a} (1-p_k^{-1} q^{-k})$,
where $\{p_1,\dots,p_{n_a}\}$ are the unstable roots of $F(q)$.

The following theorem states our main result.

\begin{thm}
\label{thm:mainresult}
Let $H(q)$ be stable and inversely stable, and factorize $F(q)$ according to~\eqref{eq:FsFa}.
The asymptotic minimizers of~\eqref{eq:Jtime} are given by
\begin{equation}
\left.
\begin{IEEEeqnarraybox}[][c]{rClCl}
\bar{A}(q) & = & \frac{1}{H(q)} \frac{F_a(q)}{F_a^*(q)} & = & \frac{\Gamma(q)D(q)}{C(q)}\frac{F_a(q)}{F_a^*(q)}, \\ 
\bar{B}(q) & = & \frac{1}{H(q)}\frac{L(q)}{\Gamma(q) F_s(q) F_a^*(q)}  & = & \frac{D(q)}{C(q)}\frac{L(q)}{F_s(q) F_a^*(q)} ,
\end{IEEEeqnarraybox}
\right.
\label{eq:ABstar} 
\end{equation}
and the attained global minimum is
\begin{IEEEeqnarray}{rCl}
J^\star & = & \abs{\frac{F_a(e^{i\omega})}{F_a^*(e^{i\omega})}}^2 \lambda_e.
\label{eq:Jstar}
\end{IEEEeqnarray}
\end{thm}

\begin{pf}
Using that $r_t$ and $e_t$ are uncorrelated and Parseval's identity, \eqref{eq:Jtime} can be written as
\begin{IEEEeqnarray}{rCl}
J = J_r + J_e ,
\label{eq:Jfreq}
\end{IEEEeqnarray}
where
\begin{IEEEeqnarray}{rCl}
J_r & = & \frac{1}{2\pi} \int_{-\pi}^{\pi} \abs{AG-B}^2 \abs{S}^2 \Phi_r \text{ d}\omega , \label{eq:Jr_st} \\
J_e & = & \frac{1}{2\pi} \int_{-\pi}^{\pi} \abs{A+KB}^2 \abs{HS}^2 \lambda_e \text{ d}\omega , \label{eq:Je_st}
\end{IEEEeqnarray}
with $\Phi_r$ the spectrum of $r_t$.
Let
\begin{IEEEeqnarray*}{rCcCl}
\tilde{S}(q) & \defeq & \frac{S(q)}{F_a(q)} & = & \frac{F_s(q)}{F(q)+K(q)L(q)} , 
\end{IEEEeqnarray*}
and re-write~\eqref{eq:Je_st} as
\begin{IEEEeqnarray*}{rCl}
J_e & = & \abs{\frac{F_a}{F_a^*}}^2 \frac{1}{2\pi} \int_{-\pi}^{\pi}  \abs{A+KB}^2 \abs{H\tilde{S}F_a^*}^2 \lambda_e \text{ d}\omega ,
\end{IEEEeqnarray*}
where $\abs{\nicefrac{F_a(e^{i\omega})}{F_a^*(e^{i\omega})}}$ is moved outside the integral since it is an all-pass filter.
Therefore, since $A(q) + K(q) B(q)$ is monic, and $H(q) \tilde{S}(q) F_a^*(q)$ is monic, stable, and inversely stable, they can be expanded as power series in $q^{-1}$ satisfying the conditions of $X(q)$ and $Z(q)$, respectively, in Proposition~\ref{thm:firstresult}. 
Then, we can use this result to conclude that $J_e$ is minimized by
\begin{IEEEeqnarray*}{rCl}
A(q) + K(q) B(q) & = & [H(q) \tilde{S}(q) F_a^*(q)]^{-1} .
\label{eq:minJe}
\yesnumber
\end{IEEEeqnarray*}
If~\eqref{eq:Jr_st} can be made zero while satisfying~\eqref{eq:minJe}, a global minimizer for $J$ has been found.
This is done by setting $\bar{A}(q)$ and $\bar{B}(q)$ according to~\eqref{eq:ABstar}, with minimum~\eqref{eq:Jstar}. $\hfill \qed$
\end{pf}

Notice that, if $F(q)$ is stable, $F_a(q) = 1 = F_a^*(q)$ and $F_s(q) = F(q)$, and \eqref{eq:ABstar} reduces to~\eqref{eq:ABstar_stable_minphase}. 
Moreover, using a similar approach to Theorem~\ref{thm:mainresult}, it is straightforward to extend this result to the case with a non-minimum phase noise model $H(q)$. 
In this case, the noise model $H(q)$ in~\eqref{eq:ABstar} will be replaced by its minimum phase equivalent. 
This corresponds to the well known result that PEM identifies an equivalent minimum phase noise model if the true noise model is non-minimum phase \cite{hjalmarsson99}.

Recovering $G(q)$ from the asymptotic minimizers of the ARX model is straightforward, as
\begin{IEEEeqnarray*}{rCl}
G(q) & = & \frac{\bar{B}(q)}{\bar{A}(q)} .
\label{eq:G_equation} \yesnumber
\end{IEEEeqnarray*}
The following corollary describes how the noise model and the variance $\lambda_e$ can be retrieved.
\begin{cor}
\label{corollary}
Let $J^\star$ be the asymptotic minimum of \eqref{eq:Jtime}, and $\bar{A}(q)$ the corresponding minimizer. 
Then, the noise model $H(q)$ and the noise variance $\lambda_e$ can be retrieved by
\begin{IEEEeqnarray*}{rCl}
H(q) & = & \frac{1}{\bar{A}(q)} \frac{\bar{A}_a(q)}{\bar{A}_a^*(q)}
\label{eq:H_equation} \yesnumber
\end{IEEEeqnarray*}
and
\begin{IEEEeqnarray*}{rCl}
\lambda_e & = & J^\star \abs{\frac{\bar{A}_a^*(e^{i\omega})}{\bar{A}_a(e^{i\omega})}}^2 ,
\label{eq:lambda_e} \yesnumber
\end{IEEEeqnarray*}
respectively. 
\end{cor}
\begin{pf}
The asymptotic minimizer $\bar{A}(q)$ can be factorized by one polynomial $F_a(q)$ with anti-stable roots and one polynomial with only stable roots corresponding to $\nicefrac{1}{H(q)F_a^*(q)}$. 
Thus, $F_a(q)$ can be retrieved as the anti-stable roots of $\bar{A}(q)$,
\begin{IEEEeqnarray*}{rCl}
F_a(q) & = & \bar{A}_a(q).
\label{eq:A_a_equation} \yesnumber
\end{IEEEeqnarray*}
Then, \eqref{eq:H_equation} follows directly from~\eqref{eq:ABstar} and~\eqref{eq:A_a_equation}, while~\eqref{eq:lambda_e} follows from~\eqref{eq:Jstar} and~\eqref{eq:A_a_equation}. $\hfill \qed$
\end{pf}

Comparing Corollary~\ref{corollary} to \eqref{eq:ABstar_stable_minphase} and \eqref{eq:Jstar_stable_minphase}, it is observed that the noise model $H(q)$ and noise variance $\lambda_e$ will be wrongly estimated if the appropriate corrections due to the unstable plant are not made. 
In particular, we observe that $\nicefrac{\bar{A}_a(q)}{\bar{A}_a^*(q)}$ is an amplifying all-pass filter.
So, without this correction factor, the magnitude of the noise model is underestimated by a constant bias, and the noise variance is overestimated.



\section{Practical Aspects}
\label{sec:var}

So far, we have only discussed consistency of the ARX model. 
For the consistency results to be valid, the ARX model has to be of infinite order.
Otherwise, the system~\eqref{eq:truemodel} is not in the model set defined by~\eqref{eq:arxmodel}, and a bias is induced by the truncation.
For estimation purposes, we consider the ARX model~\eqref{eq:arxmodel}, whose polynomials are given by
\begin{IEEEeqnarray*}{rCl+rCl}
A(q) & = & 1+\sum_{k=1}^{n_a} a_k q^{-k} , & B(q) & = & \sum_{k=1}^{n_b} b_k q^{-k} ,
\end{IEEEeqnarray*}
where the orders $n_a$ and $n_b$ are chosen large enough for these polynomials to capture the dynamics of the true system.
In that case, the bias error due to the truncation is assumed to be small.

The inherent limitation of estimating a high order model is that the estimated model will have high variance. 
However, we observe that this does not apply to the unstable poles of the ARX model, as will be illustrated with a simulation in the next section.
Thus, the estimate of $F_a(q)$ obtained from~\eqref{eq:A_a_equation} will have high accuracy in comparison to the complete high order ARX-model estimate. 
In turn, this means that the noise variance can be estimated according to~\eqref{eq:lambda_e} with high accuracy.

There is theoretical support for the observation that the unstable roots of $A(q)$ have low variance. 
For systems with the same form as~\eqref{eq:sys}, it has been shown that the variance of any unstable root will converge to a finite limit (cf. Theorem 5.1 in~\cite{Martensson09}).
However, this result is not directly applicable to the setting of this paper, 
since such variance analysis requires that the model order tend to infinity as function of the sample size $N$, similarly to the approach in~\cite{ljung&wahlberg92}.
There is no conceptual reason limiting the extension of this theorem to our setting; however, due to the technical effort required, it will be considered in a separate contribution.

\section{Examples}

Consider the plant and noise models
\begin{IEEEeqnarray*}{rCl+rCl}
G(q) & = & \frac{1q^{-1}-1.7q^{-2}}{1-2q^{-1}+2q^{-2}}, &  H(q) & = & \frac{1+0.2q^{-1}}{1-0.9q^{-1}} ,
\end{IEEEeqnarray*}
which are used to generate data according to~\eqref{eq:syscl} with the controller $K(q)=1$, where $r_t$ and $e_t$ are uncorrelated Gaussian white noise sequences with unit variance. 
Notice that the plant $G(q)$ has a pair of unstable complex poles at $1\pm i$, not shared with $H(q)$ .

We use this system for two examples. 
First, we illustrate the limit properties of the ARX model that were shown in Section~\ref{sec:arxmin}; then, we use different orders of the ARX model to illustrate the observation in Section~\ref{sec:var} regarding the variance of the estimated unstable poles.

\subsection{Limit Properties of the ARX Model}

The objective of this example is to illustrate the result obtained in Theorem~\ref{thm:mainresult}, and how Corollary~\ref{corollary} can be used to obtain estimates of $G(q)$ and $H(q)$ when $F(q)$ is unstable. 
Because these results concern the limit values, in both model order and sample size, of the estimates of $A(q)$ and $B(q)$, 
they can be more clearly illustrated if the estimation error is kept small.
Thus, to minimize the bias error due to the ARX model truncation, we choose $G(q)$ and $H(q)$ such that the coefficients of $\bar{A}(q)$ and $\bar{B}(q)$ decay quickly, which allows us to use a relatively low order ($n_a=n_b=15$).
The low model order together with a large sample size ($N=100000$) ensure that also the variance error will be small.
Finally, we are also interested in estimating the noise variance, $\lambda_e$.

The procedure is as follows. 
First, the ARX polynomials $A(q)$ and $B(q)$ are estimated by minimizing the cost function
\begin{IEEEeqnarray*}{rCl}
J & = & \frac{1}{N} \sum_{t=1}^N \left[ A(q) y_t - B(q) u_t \right]^2 ,
\end{IEEEeqnarray*}
which is a consistent estimate of~\eqref{eq:JtimeOld} for finite sample size. 
This is a least-squares problem, and yields estimates $\hat{A}(q)$ and $\hat{B}(q)$, at which the minimum $\hat{J}$ is obtained.

Then, we estimate the plant and noise model from the estimated ARX polynomials. 
Motivated by~\eqref{eq:G_equation} and~\eqref{eq:H_equation}, we calculate 
\begin{IEEEeqnarray*}{rCl+rCl}
\hat G(q) & = & \frac{\hat B(q)}{\hat A(q)}, &
\hat H(q) & = & \frac{1}{\hat A(q)}\frac{\hat{A}_a(q)}{\hat A_a^*(q)}.
\label{eq:GhatHhat}
\end{IEEEeqnarray*}

\begin{figure}
\centering
\input{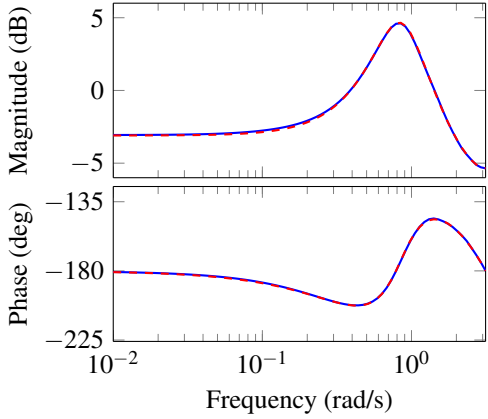}
\caption{Bode plot of $G(q)$ (red, dashed) and its estimate $\hat{G}(q)$ (blue, full).}  
\label{fig:G}                                 
\end{figure}

\begin{figure}
\centering
\vspace{4pt}
\input{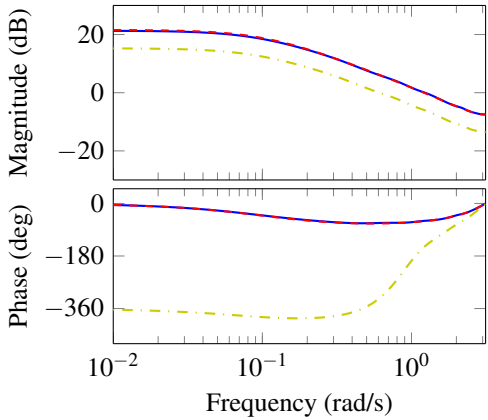}
\caption{Bode plot of $H(q)$ (red, dashed), its estimate  $\hat{H}(q)$ (blue, full), and its uncorrected estimate $\hat{A}^{-1}(q)$ (yellow, dash-dotted).}  
\label{fig:A}                                 
\end{figure}

In Fig.~\ref{fig:G} and Fig.~\ref{fig:A}, the Bode plots of $G(q)$ and $H(q)$ are shown respectively together with their corresponding estimates, and, in the case of the noise model, also the estimate without the correction for the unstable plant is shown.
Here, it is observed that the ARX model correctly captures the true system, according to~\eqref{eq:ABstar}. 
In particular, this illustrates the main result of this paper, that when the plant $G(q)$ has unstable poles and is parametrized independently of the noise model $H(q)$, a high-order ARX model is still appropriate to consistently model this system. 
However, while the standard result~\eqref{eq:ABstar_stable_minphase} still applies to consistently retrieve the plant, a consistent estimate of the noise model is obtained by using a correction factor according to~\eqref{eq:H_equation}.

Finally, we are interested in estimating the noise variance, $\lambda_e$. 
With that purpose, motivated by \eqref{eq:lambda_e}, we calculate
\begin{IEEEeqnarray*}{rCl}
\hat{\lambda}_e & = & \hat{J} \abs{\frac{\hat{A}_a^*(e^{i\omega})}{\hat{A}_a(e^{i\omega})}}^2 .
\label{eq:hatlambda_e} \yesnumber
\end{IEEEeqnarray*}
Recalling that $\lambda_e=1$, the results obtained for this example were
\begin{IEEEeqnarray*}{rCl+rCl}
\hat J & = & 3.9816, &
\hat \lambda_e & = & 0.9988 .
\end{IEEEeqnarray*}
Again, we observe how the high-order ARX model can still be used to obtain a consistent estimate of the noise variance, even in our generalized setting for unstable plants, as long as the appropriate correction~\eqref{eq:hatlambda_e} is made. 
Otherwise, taking the minimum $\hat{J}$ as estimate for the noise variance, as in~\eqref{eq:Jstar_stable_minphase}, would overestimate the noise variance by a factor of four. 

\subsection{Variance of the Estimated Unstable Poles}

As ARX models typically need to be of high order to capture the dynamics of a system with $G(q)$ and $H(q)$ given by~\eqref{eq:sys}, 
the variance of the estimated model will be large. 
Although this intrinsic limitation was not evident in the previous example, since the dynamics of the considered system can be captured with relatively low orders of $A(q)$ and $B(q)$ (and a large sample size was used), it can be made clear by letting the order of the ARX polynomials increase.
 
As the variance of the estimated $\hat A(q)$ will be large, also the variance of the estimated poles of $G(q)$ should be large, since the poles of $G(q)$ are obtained from the roots of $\hat A(q)$.
However, following the discussion in Section~\ref{sec:var}, we observe that this does not apply to the variance of the unstable poles.

To illustrate this, we perform a Monte Carlo simulation with 50 runs, where two ARX models with different orders are computed. 
The roots of $\hat{A}(q)$ are plotted in Fig. 3 and Fig. 4, for $n_a=n_b=15$ and $n_a=n_b=100$, respectively. 
Here, it is clear that the variance of the unstable poles is small relative to the stable ones,
and also that there is no apparent variance increase for the unstable poles when the number of estimated parameters increases.

\begin{figure}
\centering
\input{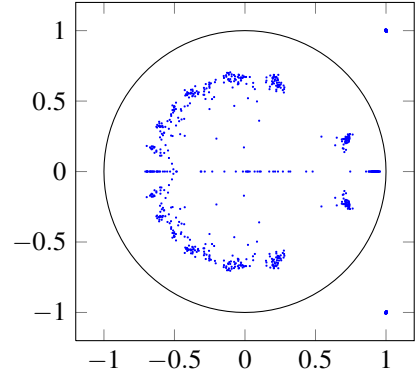}
\caption{Roots of $\hat{A}(q)$ for 50 Monte Carlo runs and $n_a=n_b=15$.}  
\label{fig:MC15}                                 
\end{figure}

\begin{figure}
\centering
\vspace{5pt}
\input{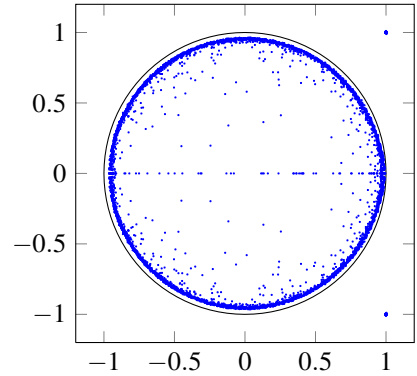}
\caption{Roots of $\hat{A}(q)$ for 50 Monte Carlo runs and $n_a=n_b=100$.}  
\label{fig:MC100}                                 
\end{figure}

\section{Discussion}

In this paper, we derived asymptotic results for the limit values of ARX models when used to model a system with an unstable plant that does not necessarily share the unstable poles with the true noise model.
High-order ARX models can still be used to model the underlying system in this situation. However, while $\nicefrac{B(q)}{A(q)}$ still captures the plant, $\nicefrac{1}{A(q)}$ no longer corresponds to the noise model $H(q)$, but will also depend on the unstable part of $F(q)$. 

This result has also implications for modeling output-error (OE) structures with high-order models. 
Typically, in this case, a high-order finite impulse response model can be estimated instead of an ARX, since $H(q)=1=A(q)$. 
However, that is only the case if $G(q)$ is stable. When $G(q)$ is unstable, a finite-order polynomial $B(q)$ is not sufficient to approximate $G(q)$ arbitrarily well. 
Thus, even if the true system is OE, an ARX model is required to asymptotically capture the system dynamics in this case.

The results in this paper also have a correspondence to those in~\cite{forssell00}. 
Therein, the authors derive modified, but asymptotically equivalent, versions of OE and BJ models that yield stable predictors in the case of unstable systems. 
The modified versions contain an additional factor $\nicefrac{F_a^*(q)}{F_a(q)}$ to be estimated in the noise model, which corresponds to the same quantity appearing in the estimate of $\nicefrac{1}{A(q)}$ here derived.

Finally, we have shown how the plant, noise model, and noise variance of a system as~\eqref{eq:truemodel} can be obtained from an estimated ARX model, in the case of an unstable plant. 
Also, we note that the high variance inherent to the high order of the ARX model does not affect the estimation of the unstable poles nor the estimation of the noise variance.

\bibliographystyle{unsrt}        
\bibliography{mybib}             

\end{document}